\documentclass[superscriptaddress, amsmath, aps, prl, longbibliography, numerical, twocolumn]{revtex4-1}

\usepackage[T1]{fontenc}
\usepackage{latexsym}
\usepackage{amssymb}
\usepackage{amsfonts}
\usepackage{amsmath}
\usepackage{amsthm}
\usepackage{multirow}
\usepackage{longtable}
\usepackage{mathtools}
\usepackage{braket}
\usepackage{siunitx}
\usepackage{xcolor}

\newcommand{\dmscr}{\delta \left\langle r_\mathrm{c}^2 \right\rangle}
\newcommand{\rskin}{\ensuremath{R_{\rm skin}}}
\newcommand{\ad}{\ensuremath{\alpha_\mathrm{D}}}

\begin{document}

\title{Charge radius of the short-lived $^{68}$Ni and correlation with the dipole polarizability}

\author{S. Kaufmann}
\affiliation{Institut f\"ur Kernphysik, Technische Universit\"at Darmstadt, D-64289 Darmstadt, Germany}
\author{J. Simonis}
\affiliation{Institut f\"ur Kernphysik and PRISMA$^+$ Cluster of
  Excellence, Johannes Gutenberg-Universit\"at Mainz, D-55128 Mainz,
  Germany}
\author{S. Bacca}
\affiliation{Institut f\"ur Kernphysik and PRISMA$^+$ Cluster of
  Excellence, Johannes Gutenberg-Universit\"at Mainz, D-55128 Mainz,
  Germany}
\affiliation{Helmholtz-Institut Mainz, Johannes Gutenberg-Universit\"at Mainz, D-55099 Mainz, Germany}
\author{J. Billowes}
\affiliation{School of Physics and Astronomy, The University of Manchester, Manchester, M13 9PL, United Kingdom}
\author{M. L. Bissell}
\affiliation{School of Physics and Astronomy, The University of Manchester, Manchester, M13 9PL, United Kingdom}
\author{K. Blaum}
\affiliation{Max-Planck-Institut f\"ur Kernphysik, D-69117 Heidelberg, Germany}
\author{B. Cheal}
\affiliation{Oliver Lodge Laboratory, Oxford Street, University of Liverpool, Liverpool, L69 7ZE, United Kingdom}
\author{R.~F.~Garcia~Ruiz}
\affiliation{School of Physics and Astronomy, The University of Manchester, Manchester, M13 9PL, United Kingdom}
\affiliation{Experimental Physics Department, CERN, CH-1211 Geneva 23, Switzerland}
\author{W. Gins}
\affiliation{KU Leuven, Instituut voor Kern- en Stralingsfysica, B-3001 Leuven, Belgium}
\author{C. Gorges}
\affiliation{Institut f\"ur Kernphysik, Technische Universit\"at Darmstadt, D-64289 Darmstadt, Germany}
\author{G. Hagen}
\affiliation{Physics Division, Oak Ridge National Laboratory, Oak
  Ridge, Tennessee 37831, USA and Department of Physics and Astronomy, University of Tennessee, Knoxville, TN 37996, USA}
\author{H. Heylen}
\affiliation{Max-Planck-Institut f\"ur Kernphysik, D-69117 Heidelberg, Germany}
\affiliation{Experimental Physics Department, CERN, CH-1211 Geneva 23, Switzerland}
\author{A. Kanellakopoulos}
\affiliation{KU Leuven, Instituut voor Kern- en Stralingsfysica, B-3001 Leuven, Belgium}
\author{S. Malbrunot-Ettenauer}
\affiliation{Experimental Physics Department, CERN, CH-1211 Geneva 23, Switzerland}
\author{M.~Miorelli}
\affiliation{TRIUMF, 4004 Wesbrook Mall, Vancouver, British Columbia,
  V6T 2A3, Canada}
\author{R. Neugart}
\affiliation{Max-Planck-Institut f\"ur Kernphysik, D-69117 Heidelberg, Germany}
\affiliation{Institut f\"ur Kernchemie, Johannes Gutenberg-Universit\"at Mainz, D-55128 Mainz, Germany}
\author{G. Neyens}
\affiliation{Experimental Physics Department, CERN, CH-1211 Geneva 23, Switzerland}
\affiliation{KU Leuven, Instituut voor Kern- en Stralingsfysica, B-3001 Leuven, Belgium}
\author{W. N\"ortersh\"auser}
\email{wnoertershaeuser@ikp.tu-darmstadt.de}
\affiliation{Institut f\"ur Kernphysik, Technische Universit\"at Darmstadt, D-64289 Darmstadt, Germany}
\author{R. S\'anchez}
\affiliation{GSI Helmholtzzentrum f\"ur Schwerionenforschung, D-64291 Darmstadt, Germany}
\author{S. Sailer}
\affiliation{Technische Universit\"at M\"unchen, D-80333 M\"unchen, Germany}
\author{A. Schwenk}
\affiliation{Institut f\"ur Kernphysik, Technische Universit\"at Darmstadt, D-64289 Darmstadt, Germany}
\affiliation{Max-Planck-Institut f\"ur Kernphysik, D-69117 Heidelberg, Germany}
\affiliation{ExtreMe Matter Institute EMMI, GSI Helmholtzzentrum f\"ur Schwerionenforschung GmbH, D-64291 Darmstadt, Germany}
\author{T. Ratajczyk}
\affiliation{Institut f\"ur Kernphysik, Technische Universit\"at Darmstadt, D-64289 Darmstadt, Germany}
\author{L. V. Rodr\'iguez}
\affiliation{Institut de Physique Nucl\'eaire, CNRS-IN2P3, Universit\'e Paris-Sud, Universit\'e Paris-Saclay, 91406 Orsay, France}
\author{L. Wehner}
\affiliation{Institut f\"ur Kernchemie, Universit\"at Mainz, D-55128 Mainz, Germany}
\author{C. Wraith}
\affiliation{Oliver Lodge Laboratory, Oxford Street, University of Liverpool, Liverpool, L69 7ZE, United Kingdom}
\author{L. Xie}
\affiliation{School of Physics and Astronomy, The University of Manchester, Manchester, M13 9PL, United Kingdom}
\author{Z. Y. Xu}
\affiliation{KU Leuven, Instituut voor Kern- en Stralingsfysica, B-3001 Leuven, Belgium}
\author{X. F. Yang}
\affiliation{KU Leuven, Instituut voor Kern- en Stralingsfysica, B-3001 Leuven, Belgium}
\affiliation{School of Physics and State Key Laboratory of Nuclear Physics and Technology, Peking University, Beijing 100871, China}
\author{D. T. Yordanov}
\affiliation{Institut de Physique Nucl\'eaire, CNRS-IN2P3, Universit\'e Paris-Sud, Universit\'e Paris-Saclay, 91406 Orsay, France}

\begin{abstract}

We present the first laser spectroscopic measurement of the neutron-rich nucleus $^{68}$Ni  at the  \mbox{$N=40$} subshell closure and extract its nuclear charge radius. 
Since this is the only short-lived isotope for which the dipole polarizability $\ad$ has been measured, the combination of these observables provides a benchmark for nuclear structure theory. We compare them to novel coupled-cluster calculations based on different chiral two- and three-nucleon interactions, for which a strong correlation between the charge radius and dipole polarizability is observed, similar to the stable nucleus $^{48}$Ca. Three-particle--three-hole correlations in coupled-cluster theory substantially improve the description of the experimental data, which allows to constrain the neutron radius and neutron skin of $^{68}$Ni.
\end{abstract}

\maketitle

\section{Introduction}

The nuclear equation of state (EOS) plays a key role in supernova explosions and compact object mergers. In fact, the gravitational wave signal from the neutron star merger GW170817 has recently lead to constraints on the EOS of neutron-rich matter \cite{ligo18}, which is consistent with our knowledge of nuclear physics. 
While the EOS of symmetric nuclear matter is well constrained around saturation density \cite{danielewicz2002}, the properties of neutron-rich matter are still rather uncertain. These properties are encoded in the nuclear symmetry energy $S(n)$ as a function of density $n$ and the slope parameter $L = 3 n_0 \partial S(n_0) / \partial n$ at saturation density $n_0$.
Studies on atomic nuclei can provide information on the $L$ parameter \cite{Tsan12esymm} through a nuclide's neutron-skin thickness 
$\rskin = R_\mathrm{n} - R_\mathrm{p}$ defined as the difference between the point-neutron and point-proton radii. 
The neutron skin is a consequence of the competition between the surface tension and the pressure of neutron matter, which is determined by the $L$ parameter.
In the heavy nucleus $^{208}$Pb energy density functional (EDF) calculations confirmed this strong correlation between $\rskin$ and $L$ with a correlation coefficient of 0.979. This allows one to further constrain $L$ based on $\rskin$ \cite{rocamaza11}. 
Unfortunately, the direct measurement of $\rskin$ is experimentally very challenging. In recent measurements it was extracted by its correlation to the dipole polarizability $\ad$, which can be explored, e.g., with proton inelastic scattering, as it was the case for $^{48}$Ca \cite{birkhan17}, $^{120}$Sn \cite{hashimoto15} and $^{208}$Pb \cite{tamii11}. 
Here mostly EDFs were used to extract the neutron skin from the dipole polarizability, but in the case of $^{48}$Ca the neutron skin was predicted from first principles coupled-cluster calculations to be surprisingly small, only 0.12--0.15\,fm \cite{hagen15}. These ab initio calculations starting from two- and three-nucleon interactions based on chiral effective field theory (EFT) \cite{epelbaum09, machleidt11, Hamm13RMP} further revealed a correlation between the charge radius and the dipole polarizability, which was predicted to be in the range 2.19--2.60\,fm$^3$. 
Recent measurements of $^{48}$Ca by Birkhan \textit{et al.} \cite{birkhan17} yielded a dipole polarizability of $\ad=2.07(22)$\,fm$^3$ in agreement with the chiral EFT predictions. 
The only short-lived nucleus for which $\ad$ has been experimentally determined is $^{68}$Ni, using Coulomb excitation in inverse kinematics. The pygmy and the giant dipole resonances were observed and $\ad$ was extracted \cite{rossi13}.
In this Letter, we focus on the charge radius of $^{68}$Ni, determined by collinear laser spectroscopy. 
It is the first laser spectroscopy result on a short-lived nickel isotope, since access to this element at ISOL facilities is limited due to the slow release from the target. On the theory side, we report on the first coupled-cluster calculation including triples of $R_\mathrm{c}$ and \ad\ of $^{68}$Ni based on chiral EFT interactions, being now the heaviest system for which this has been achieved. 
We study the  correlation of the charge radius with the dipole polarizability in novel ab initio calculations including triples contributions. 
The measured charge radius in combination with the experimental dipole polarizability enables the first test of this correlation in a neutron-rich medium-mass nucleus.

\section{Experiment}

\begin{figure}[t]
	\centering
		\includegraphics[width=.4\textwidth]{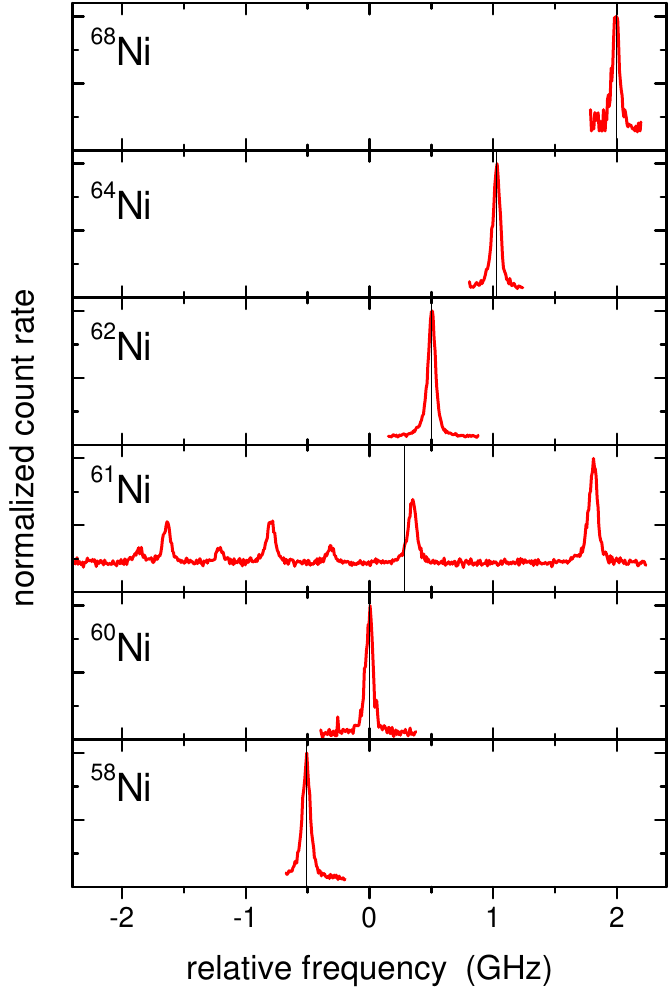}
	\caption{Spectra of the stable isotopes $^{58,60,61,62,64}$Ni and the radioactive $^{68}$Ni with their center frequency 
	indicated with a vertical line. The count rate is normalized for each isotope, and the frequency is given relative to the center frequency of the reference isotope $^{60}$Ni.}
	\label{fig:shift_spectra}
\end{figure}

Nickel isotopes were produced at ISOLDE/CERN 
using proton pulses at an energy of 1.4\,GeV to cause fragmentation, spallation, and fission inside a uranium carbide target. The target was heated beyond standard operation temperatures up to $\sim 2200\,^{\circ}$C to enhance the release of chemically reactive nickel isotopes that have generally quite long release times from the target. Nickel atoms were then ionized by resonant laser ionization using RILIS \cite{marsh14} and accelerated towards the high-resolution mass separator on ground potential by applying an  electrostatic potential of approximately 30\,kV and 40\,kV to the ion source in a first and a second beamtime, respectively.  
The mass separated ions were injected into the radio-frequency ion beam cooler and buncher ISCOOL \cite{franberg08} and accumulated for typically $10-100$\,ms. 
Extracted ion bunches of 5\,$\mu$s duration were transported to the collinear laser spectroscopy beam line COLLAPS where the ions were superimposed with a co-propagating laser beam. 
Potassium vapor in a charge-exchange cell 
\cite{muller83, klose12} was used to neutralize the ions. Various excited states of the nickel atoms were populated within this non-resonant process, among them the metastable $3d^9\,4s\,^{3}D_3$ level \cite{ryder15} that served as the starting point for laser spectroscopy, performed in the 352.45\,nm transition to the $3d^9\,4p\,^{3}P_2$ level. 
Fluorescence photons from spontaneous emission were detected by four photomultiplier tubes and the individual events were recorded with a new time-resolving data acquisition system. The laser light was produced using a frequency-doubled single-mode continuous-wave titanium sapphire laser stabilized on a high-resolution wavemeter, which was calibrated regularly with a stabilized helium-neon laser. 
Typical spectra of the isotopes $^{58,60,61,62,64,68}$Ni are shown in Fig.~\ref{fig:shift_spectra}. All isotopes were measured alternating with the reference isotope $^{60}$Ni to compensate for any remaining long-term drifts in the ion velocity or the laser frequency.

\section{Analysis}

\begin{figure}[t]
	\centering
		\includegraphics[width=\columnwidth]{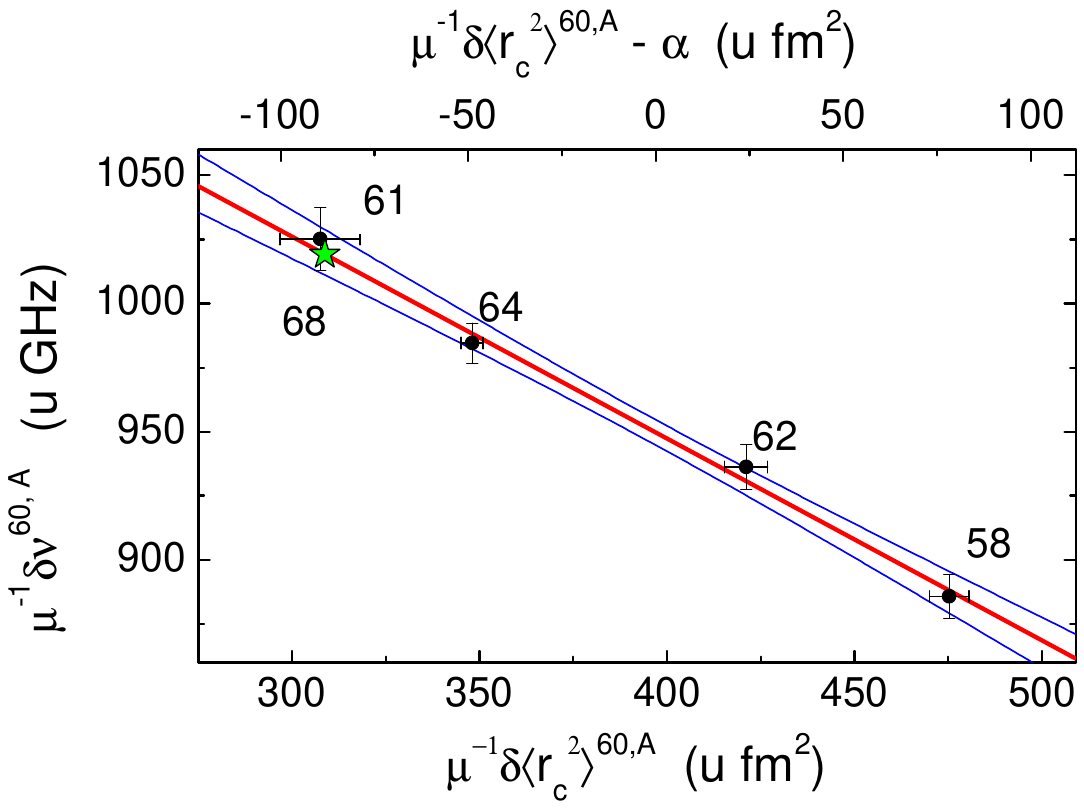}
	\caption{King-plot analysis of the stable isotopes. 
	The red line is a straight line fit to the black data points taking $x$- and $y$-errors into account and the $1\sigma$-confidence interval is shown as a blue  solid line. The top axis is shifted by $\alpha\!=\!397$\,u\,fm$^2$ to remove the correlation between $F$ and $M$. The green star shows the position of $^{68}$Ni, from which the charge radius is extracted.}
	\label{fig:kingFit}
\end{figure}

Isotope shifts $\delta \nu^{60,A} =  \nu^{A} - \nu^{60}$ for the stable isotopes $^{58,61,62,64}$Ni and the radioactive $^{68}$Ni were calculated from their respective center frequency $\nu^{A}$ with respect to the center frequency $\nu^{60}$ of the reference isotope $^{60}$Ni. Results are listed in Table \ref{tab:shifts}. Isotope shifts are related to differences in mean-square charge radii $\dmscr^{60,A}$ via a field shift factor $F$ and a mass shift factor $M$ according to
\begin{equation}
\delta \nu^{60,A} =  
\mu \cdot M + F \cdot \dmscr^{60,A},
\label{eq:shift}
\end{equation}
with $\mu = (m_A - m_{60}) / (m_A \cdot m_{60})$ and $m_A$ being the respective atomic masses. A King-fit analysis was performed using the  procedure described in \cite{hammen18} and based on the known rms charge radii of the stable nickel isotopes extracted from the combined analysis of muonic atom data and elastic electron scattering provided in \cite{fricke04}. 
An $x$-axis offset of $\alpha\!=\!397$\,u\,fm$^2$ was used to remove the correlation between $M$ and $F$ and the result of the fit is depicted in Fig.~\ref{fig:kingFit}. 
The intercept of the line with the new $y$-axis provides $M_{\alpha\!=\!397}\!=\!949(4)$\,GHz\,u [corresponding to a mass shift parameter of $M\!=\!1262(32)$\,GHz\,u] and a field shift parameter of $F=-788(82)\,\mathrm{MHz/fm}^2$.  
The isotope shift of $^{68}$Ni puts it on the line at the position indicated by a star and the change in the ms charge radius can be calculated with respect to $^{60}$Ni according to   
\begin{equation}
\mu^{-1} \, \dmscr^{60,A}  = \left( \mu^{-1} \, \delta \nu^{60,A} -  M_{\alpha}\right) / F + \alpha .
\label{eq:shiftAlpha}
\end{equation}
The results are listed in Table \ref{tab:shifts}. With the rms charge radius of 
$R_\mathrm{c}(^{60}\mathrm{Ni})\!=\!3.806(2)$\,fm taken from \cite{fricke04}, the charge radius of $^{68}$Ni is obtained as 
$R_\mathrm{c}(^{68}\mathrm{Ni})\!=\!3.887(3)$\,fm.

\begin{table}[b]
	\centering
		\caption{Measured isotope shifts $\delta \nu^{60,A}$ relative to $^{60}$Ni with statistical uncertainties in parentheses and systematic uncertainties in square brackets. The statistical uncertainty includes variations between the two beamtimes that are partially of systematic but uncorrelated origin and change statistically from isotope to isotope, while the systematic uncertainty is restricted to the correlated uncertainty caused by the high-voltage measurement. The extracted change in ms charge radius $\dmscr^{60,A}$ and the total charge radii $R_\mathrm{c}$ are listed with the total uncertainties.}
\begin{tabular}{l p{0.2cm} S[table-number-alignment = right] l p{0.1cm} S S}
\hline \hline
$A$ & & \multicolumn{2}{c}{$\delta \nu^{60,A}$/MHz} & & {$\dmscr^{60,A}$/fm$^2$} & {$R_\mathrm{c}$/fm} \\
\hline 
58 & & -509.1(25) & \,\,\,\,\,[42] & & -0.275(7) & 3.770(2) \\
60 & & 0.0 & & & 0.0 & 3.806(2) \\
61 & & 280.8(27) & \,\,\,\,\,[20] & & 0.083(5) & 3.817(2) \\
62 & & 503.9(25) & \,\,\,\,\,[39] & & 0.223(5) & 3.835(2) \\
64 & & 1027.2(25) & \,\,\,\,\,[77] & & 0.368(9) & 3.854(2) \\
68 & & 1992.3(27) & \,\,\,\,\,[147] & & 0.620(21) & 3.887(3) \\
\hline
		\end{tabular}
	\label{tab:shifts}
\end{table}

\section{Discussion}

The extracted $R_\mathrm{c}$ can be used to benchmark theoretical calculations, to test and expand their reliability and predictive power away from stable nuclei. 
First principle calculations were recently performed for $^{48}$Ca~\cite{hagen15}, which lead to an improved understanding of the neutron and proton distributions in nuclei, as well as their difference encoded in $\rskin$. 
The observed correlation between the dipole polarizability $\ad$ and the rms charge radius of $^{48}$Ca allowed to narrow down constraints on the dipole polarizability, $\ad=2.19-2.60\,\mathrm{fm}^3$, and on the neutron skin, $\rskin=0.12-0.15\,\mathrm{fm}$. The latter was found to be considerably smaller than previously thought \cite{hagen15}. The recent Darmstadt-Osaka experimental determination of the dipole polarizability $\ad=2.07(22)\,\mathrm{fm}^3$ \cite{birkhan17} is indeed in good agreement with the theoretical predictions.  
Subsequently, new calculations have included higher-order coupled-cluster correlations~\cite{miorelli18}, so-called linearized 3 particles--3 holes ($3p$--$3h$) correlations. This leads to a reduction of the dipole polarizability and to an improved agreement with the experimental data for $^{48}$Ca, while the charge radius is found to not depend sensitively on $3p$--$3h$ correlations~\cite{Simonis}. 

\begin{figure}[t]
	\centering
		\includegraphics[width=\columnwidth]{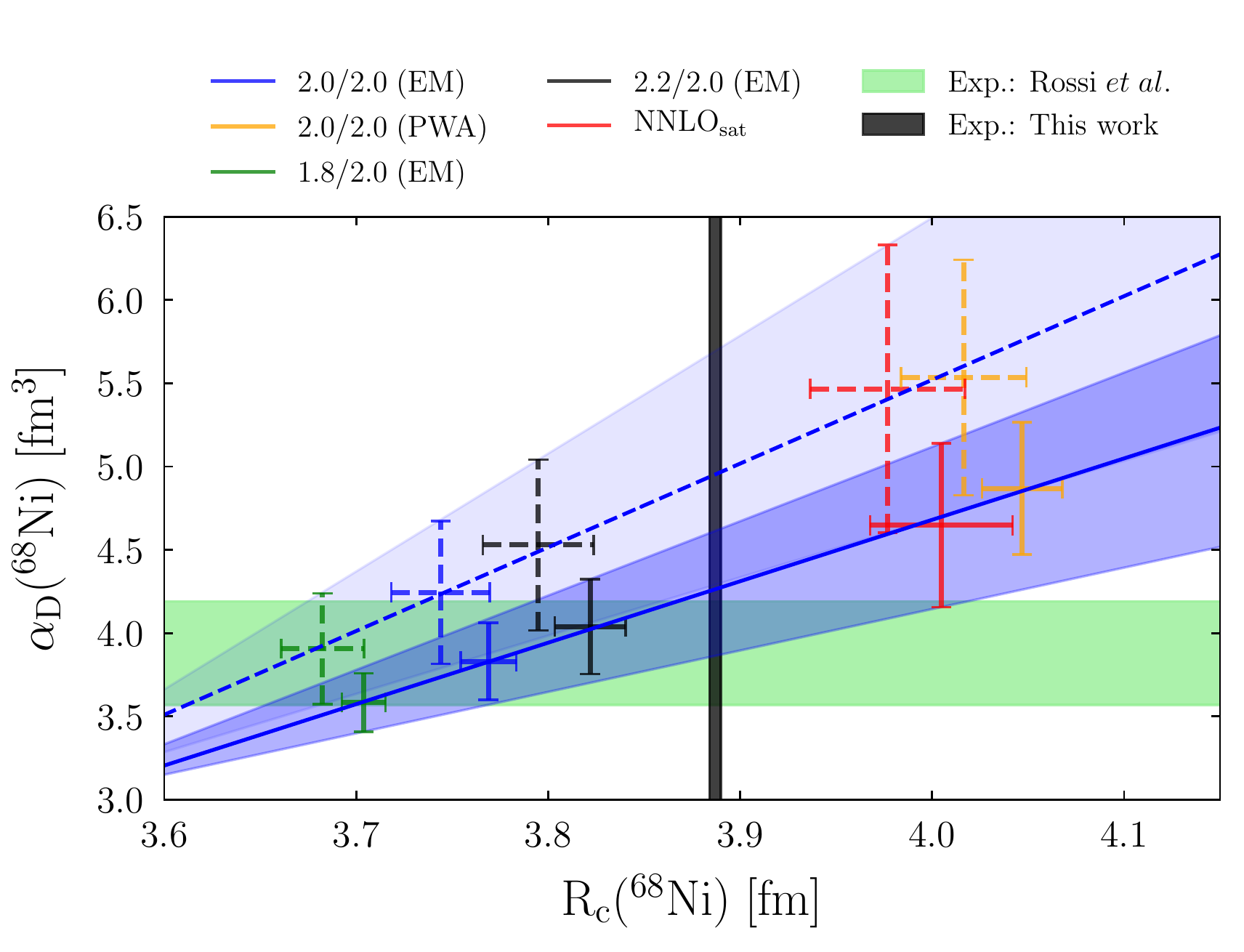}
	\caption{Theoretical results for the rms charge radius $R_\mathrm{c}$ and the dipole polarizability $\ad$ of $^{68}$Ni in comparison with experiment. The horizontal bar represents $\ad$ measured by Rossi \textit{et al.}~\cite{rossi13} and refined by Roca-Maza \textit{et al.}~\cite{rocamaza15} to $\ad\!=\!3.88(31)\,\mathrm{fm}^3$. The vertical bar represents the rms charge radius $R_\mathrm{c}\!=\!3.887(3)$\,fm determined in this work. The blue lines and bands are fits to the theoretical results for the different chiral NN and 3N interactions (crosses, see text for details) based on the coupled-cluster calculations with singles and doubles excitations (dashed line and light blue band) and when $3p$--$3h$ correlations are included (solid line and darker blue band). The widths of the bands are chosen to include the full error bars of the individual calculations.}
	\label{fig:dipolarizability}
\end{figure}

Coupled-cluster calculations of $\ad$ based on chiral EFT interactions, initiated in Refs.~\cite{bacca2013,bacca2014,hagen15,miorelli2016,miorelli18,Simonis}, have progressed towards heavier, more complex nuclei and have now reached the short-lived $^{68}$Ni. Contrary to the stable isotopes, for which inelastic proton scattering was used to experimentally access the dipole polarizability, $\ad$ of $^{68}\mathrm{Ni}$ was determined using Coulomb excitation in inverse kinematics by measuring the invariant mass in the one- and two-neutron decay channels~\cite{rossi13}. This result, subsequently refined in Ref.~\cite{rocamaza15}, is shown together with
our first experimental determination of $R_\mathrm{c}$ in Fig.~\ref{fig:dipolarizability}. Figure~\ref{fig:dipolarizability} also shows our theoretical results using four different chiral nucleon-nucleon (NN) and three-nucleon (3N) interactions from Ref.~\cite{Hebeler2011} (with the same labeling used here: 1.8/2.0, 2.0/2.0, 2.2/2.0 (EM) and 2.0/2.0 (PWA)) as well as the NNLO$_\mathrm{sat}$ interaction from Ref.~\cite{ekstrom2015}. The Hamiltonians of Ref.~\cite{Hebeler2011} are based on a chiral N$^3$LO NN potential evolved to low resolution using the similarity renormalization group combined with N$^2$LO 3N interactions fit to the $^3$H binding energy and the $^4$He charge radius. These interactions have been successfully used to study the structure of medium-mass nuclei up to $^{100}$Sn (see, e.g., Refs.~\cite{Simo16unc,Simo17SatFinNuc,Morr17Tin,Holt2019}). 

Figure~\ref{fig:dipolarizability} shows two sets of coupled-cluster calculations: one with singles and doubles correlations (dashed points, line and light blue band) and another one where the leading $3p$--$3h$ correlations are included (solid points and line with darker blue band). We observe that triples corrections lead to a sizable reduction of $\ad$ (from $8\%$ for the softest NN+3N interactions 1.8/2.0 (EM) to $15\%$ for the hardest NNLO$_\mathrm{sat}$ interaction), while $R_c$ is changed only mildly (maximally $0.7\%$ for the hardest interaction). Each theoretical point is shown with a corresponding estimate of the theoretical uncertainty, which includes both the residual model-space dependence and the coupled-cluster truncation error, following the protocol explained in Ref.~\cite{Simonis}. As expected, the uncertainties are smaller (larger) for soft (hard) interactions. For completeness, the charge radius $R_\mathrm{c}$ is obtained from the point-proton radius $R_\mathrm{p}$ by
\begin{equation}
R_\mathrm{c}^2=R_\mathrm{p}^2 + \langle r_\mathrm{p}^2 \rangle + (N/Z)\langle r_\mathrm{n}^2 \rangle + (3/4M^2) + \langle r^2 \rangle_\mathrm{so} \,,
\end{equation}
where $\langle r_\mathrm{p}^2 \rangle\!=\!0.7080(32)$\,fm$^2$ \cite{codata18} and $\langle r_\mathrm{n}^2 \rangle\!=\!-0.117(4)$\,fm$^2$ \cite{kopecky97} are the rms charge radii of a proton and a neutron, respectively, $(3/4M^2)\!=\!0.033\,\mathrm{fm}^2$ \cite{friar97} is the relativistic Darwin-Foldy correction and $\langle r^2 \rangle_\mathrm{so}$
is the spin-orbit correction for $^{68}$Ni, which we calculate consistently for each Hamiltonian.

It is interesting to note that the behavior is very similar to that observed for the stable nucleus $^{48}$Ca~\cite{hagen15,birkhan17,miorelli18,Simonis}. The theoretical results exhibit a clear correlation between the dipole polarizability and the charge radius both at the singles-and-doubles and triples excitations level. The inclusion of triples excitations, however, alters the slope of this correlation.  
In Fig.~\ref{fig:dipolarizability}, the correlation is highlighted by the linear fit to the calculations and the corresponding blue uncertainty bands, which are chosen to include the full error bars of the individual calculations. Most notably, for the results with singles and doubles excitations, the band does not overlap with the intersection region of the measured $R_c$ and $\ad$. When including  $3p$--$3h$ correlations, the theoretical band nicely overlaps with the experimental constraints. This shows that $3p$--$3h$ correlations are not negligible, and that state-of-the-art coupled-cluster computations are reliable for this first test of the charge radius and $\ad$ of the neutron-rich nucleus $^{68}$Ni.

Compared to the results $\ad = 3.60$ fm$^3$ obtained recently by Raimondi and Barbieri~\cite{Raimondi:2018mtv} with the self-consistent Green's function method using the NNLO$_\mathrm{sat}$ interaction, we obtain a considerably larger value of $\ad = 4.65(49)$\,fm$^3$ using the same interaction. The reason of the discrepancy could either be due to the different method used, or more likely related to the larger $\hbar \Omega$ value in Ref.~\cite{Raimondi:2018mtv}. The central value of our coupled-cluster results are obtained with $\hbar\Omega=12$\,MeV, which shows a very nice convergence pattern as a function of the model-space size; while we have observed that larger $\hbar\Omega$ values exhibit a much slower convergence.

\begin{table*}[t]
\caption{Coupled-cluster results including $3p$--$3h$ corrections for $\ad$ in fm$^3$ and $R_{\rm p}$, $R_{\rm n}$, $\rskin$, $R_{\rm c}$ in~fm for $^{68}$Ni for the different chiral NN and 3N interactions studied~\cite{Hebeler2011,ekstrom2015}. The central values are for the optimal harmonic-oscillator frequency $\hbar\Omega =12$ MeV.}
\label{tab:CC}
\renewcommand{\tabcolsep}{1.9mm}
\begin{tabular}{l|lllll}
\hline
Hamiltonian & $\ad$ & $R_\mathrm{p}$ & $R_\mathrm{n}$ & $\rskin$ & $R_\mathrm{c}$\\
\hline
1.8/2.0 (EM) & 3.58(18) & 3.62(1) & 3.82(1) & 0.201(1) & 3.70(1) \\
2.0/2.0 (EM) & 3.83(23) & 3.69(2) & 3.89(2) & 0.202(3) & 3.77(1) \\
2.2/2.0 (EM) & 4.04(28) & 3.74(2) & 3.94(2) & 0.203(4) & 3.82(2) \\
2.0/2.0 (PWA) & 4.87(40) & 3.97(2) & 4.17(3) & 0.204(8) & 4.05(2) \\
NNLO$_{\rm sat}$ & 4.65(49) & 3.93(4) & 4.11(5) & 0.183(8) & 4.00(4) \\
\hline
\end{tabular}
\end{table*}

In addition to our ab initio calculations, the dipole polarizability of $^{68}$Ni was also studied with nuclear EDFs, which suggested that $\ad$ is strongly correlated with the neutron skin and this correlation is even stronger when information on the symmetry energy is taken into account~\cite{rocamaza15}. This led to a prediction of the neutron skin $\rskin\!=\!0.16(4)$\,fm~\cite{rocamaza15} of $^{68}$Ni. In Table~\ref{tab:CC} we list the coupled-cluster results including $3p$--$3h$ corrections for the optimal $\hbar\Omega$ for the different chiral NN and 3N interactions studied. Taking as best interactions for the correlation plot, Fig.~\ref{fig:dipolarizability}, the ones closest to the intersection region, 2.0/2.0 (EM), 2.2/2.0 (EM), and NNLO$_{\rm sat}$, we predict in Table\,\ref{tab:CC} a range for the point-neutron radius $R_n=3.9-4.1$\,fm of $^{68}$Ni and its neutron skin $\rskin = 0.18-0.20$\,fm, in very good agreement with the EDF correlation prediction of Ref.~\cite{rocamaza15}.

\section{Summary}

We have presented the first measurement of the isotope shift of the neutron-rich $^{68}$Ni isotope by collinear laser spectroscopy. This enabled the extraction of the rms charge radius to $R_\mathrm{c}\!=\!3.887(3)$\,fm based on a King-plot analysis and the known charge radii of the stable nickel isotopes. 
This radius is used to benchmark coupled-cluster calculations including novel triples corrections for a range of chiral NN and 3N interactions.
A strong correlation between the charge radius and the dipole polarizability is shown by the theoretical calculations. Our results including the leading $3p$--$3h$ contributions agree 
much better with the experimental data compared to the case when triples corrections are neglected. In particular the theoretical correlation band intersects nicely with the measured $R_c$ and $\ad$ bands. This correlation combined with coupled-cluster calculations of the point-neutron radius and neutron skin of $^{68}$Ni allows these to be constrained to $R_n=3.9-4.1$\,fm and $\rskin = 0.18-0.20$\,fm.

\section{Acknowledgments}

We acknowledge the support of the ISOLDE Collaboration and technical teams and funding from the European Union's Horizon 2020 programme under grant agreement no.\ 654002. This work was supported by the Deutsche Forschungsgemeinschaft (DFG, German Research Foundation) -- Projektnummer 279384907 -- SFB 1245, the Collaborative Research Center [The Low-Energy Frontier of the Standard Model (SFB 1044)], the Cluster of Excellence ``Precision Physics, Fundamental Interactions, and Structure of Matter'' (PRISMA$^+$ EXC 2118/1) funded by DFG within the German Excellence Strategy -- Projektnummer 39083149 --, the BMBF under Contract Nos.~05P18RDCIA and 05P19RDFN1, the FWO (Belgium), GOA 15/010 from KU Leuven, the Office of Nuclear Physics, U.S.~Department of Energy, under Grants DESC0018223 (NUCLEI SciDAC-4 collaboration), and by the Field Work Proposal ERKBP72 at Oak Ridge National Laboratory (ORNL). Computer time was provided by the Innovative and Novel Computational Impact on Theory and Experiment (INCITE) program. The new calculations presented in this work were also performed on ``Mogon II'' at Johannes Gutenberg-Universit\"{a}t in Mainz. This work was also supported by consolidated grants from STFC (UK) - ST/L005670/1, ST/L005794/1, ST/P004423/1, and ST/P004598/1 

\bibliographystyle{apsrev4-1}

\begin{thebibliography}{35}%
\makeatletter
\providecommand \@ifxundefined [1]{%
 \@ifx{#1\undefined}
}%
\providecommand \@ifnum [1]{%
 \ifnum #1\expandafter \@firstoftwo
 \else \expandafter \@secondoftwo
 \fi
}%
\providecommand \@ifx [1]{%
 \ifx #1\expandafter \@firstoftwo
 \else \expandafter \@secondoftwo
 \fi
}%
\providecommand \natexlab [1]{#1}%
\providecommand \enquote  [1]{``#1''}%
\providecommand \bibnamefont  [1]{#1}%
\providecommand \bibfnamefont [1]{#1}%
\providecommand \citenamefont [1]{#1}%
\providecommand \href@noop [0]{\@secondoftwo}%
\providecommand \href [0]{\begingroup \@sanitize@url \@href}%
\providecommand \@href[1]{\@@startlink{#1}\@@href}%
\providecommand \@@href[1]{\endgroup#1\@@endlink}%
\providecommand \@sanitize@url [0]{\catcode `\\12\catcode `\$12\catcode
  `\&12\catcode `\#12\catcode `\^12\catcode `\_12\catcode `\%12\relax}%
\providecommand \@@startlink[1]{}%
\providecommand \@@endlink[0]{}%
\providecommand \url  [0]{\begingroup\@sanitize@url \@url }%
\providecommand \@url [1]{\endgroup\@href {#1}{\urlprefix }}%
\providecommand \urlprefix  [0]{URL }%
\providecommand \Eprint [0]{\href }%
\providecommand \doibase [0]{http://dx.doi.org/}%
\providecommand \selectlanguage [0]{\@gobble}%
\providecommand \bibinfo  [0]{\@secondoftwo}%
\providecommand \bibfield  [0]{\@secondoftwo}%
\providecommand \translation [1]{[#1]}%
\providecommand \BibitemOpen [0]{}%
\providecommand \bibitemStop [0]{}%
\providecommand \bibitemNoStop [0]{.\EOS\space}%
\providecommand \EOS [0]{\spacefactor3000\relax}%
\providecommand \BibitemShut  [1]{\csname bibitem#1\endcsname}%
\let\auto@bib@innerbib\@empty
\bibitem [{\citenamefont {Abbott}\ \emph {et~al.}(2018)\citenamefont {Abbott},
  \citenamefont {Abbott}, \citenamefont {Abbott}, \citenamefont {Acernese},
  \citenamefont {Ackley}, \citenamefont {Adams}, \citenamefont {Adams},
  \citenamefont {Addesso}, \citenamefont {Adhikari}, \citenamefont {Adya} \emph
  {et~al.}}]{ligo18}%
  \BibitemOpen
  \bibfield  {author} {\bibinfo {author} {\bibfnamefont {B.~P.}\ \bibnamefont
  {Abbott}}, \bibinfo {author} {\bibfnamefont {R.}~\bibnamefont {Abbott}},
  \bibinfo {author} {\bibfnamefont {T.~D.}\ \bibnamefont {Abbott}}, \bibinfo
  {author} {\bibfnamefont {F.}~\bibnamefont {Acernese}}, \bibinfo {author}
  {\bibfnamefont {K.}~\bibnamefont {Ackley}}, \bibinfo {author} {\bibfnamefont
  {C.}~\bibnamefont {Adams}}, \bibinfo {author} {\bibfnamefont
  {T.}~\bibnamefont {Adams}}, \bibinfo {author} {\bibfnamefont
  {P.}~\bibnamefont {Addesso}}, \bibinfo {author} {\bibfnamefont {R.~X.}\
  \bibnamefont {Adhikari}}, \bibinfo {author} {\bibfnamefont {V.~B.}\
  \bibnamefont {Adya}},  \emph {et~al.} (\bibinfo {collaboration} {The LIGO
  Scientific Collaboration and the Virgo Collaboration}),\ }\href {\doibase
  10.1103/PhysRevLett.121.161101} {\bibfield  {journal} {\bibinfo  {journal}
  {Phys. Rev. Lett.}\ }\textbf {\bibinfo {volume} {121}},\ \bibinfo {pages}
  {161101} (\bibinfo {year} {2018})}\BibitemShut {NoStop}%
\bibitem [{\citenamefont {Danielewicz}\ \emph {et~al.}(2002)\citenamefont
  {Danielewicz}, \citenamefont {Lacey},\ and\ \citenamefont
  {Lynch}}]{danielewicz2002}%
  \BibitemOpen
  \bibfield  {author} {\bibinfo {author} {\bibfnamefont {P.}~\bibnamefont
  {Danielewicz}}, \bibinfo {author} {\bibfnamefont {R.}~\bibnamefont {Lacey}},
  \ and\ \bibinfo {author} {\bibfnamefont {W.~G.}\ \bibnamefont {Lynch}},\
  }\href {\doibase 10.1126/science.1078070} {\bibfield  {journal} {\bibinfo
  {journal} {Science}\ }\textbf {\bibinfo {volume} {298}},\ \bibinfo {pages}
  {1592} (\bibinfo {year} {2002})}\BibitemShut {NoStop}%
\bibitem [{\citenamefont {Tsang}\ \emph {et~al.}(2012)\citenamefont {Tsang},
  \citenamefont {Stone}, \citenamefont {Camera}, \citenamefont {Danielewicz},
  \citenamefont {Gandolfi}, \citenamefont {Hebeler}, \citenamefont {Horowitz},
  \citenamefont {Lee}, \citenamefont {Lynch}, \citenamefont {Kohley} \emph
  {et~al.}}]{Tsan12esymm}%
  \BibitemOpen
  \bibfield  {author} {\bibinfo {author} {\bibfnamefont {M.~B.}\ \bibnamefont
  {Tsang}}, \bibinfo {author} {\bibfnamefont {J.~R.}\ \bibnamefont {Stone}},
  \bibinfo {author} {\bibfnamefont {F.}~\bibnamefont {Camera}}, \bibinfo
  {author} {\bibfnamefont {P.}~\bibnamefont {Danielewicz}}, \bibinfo {author}
  {\bibfnamefont {S.}~\bibnamefont {Gandolfi}}, \bibinfo {author}
  {\bibfnamefont {K.}~\bibnamefont {Hebeler}}, \bibinfo {author} {\bibfnamefont
  {C.~J.}\ \bibnamefont {Horowitz}}, \bibinfo {author} {\bibfnamefont
  {J.}~\bibnamefont {Lee}}, \bibinfo {author} {\bibfnamefont {W.~G.}\
  \bibnamefont {Lynch}}, \bibinfo {author} {\bibfnamefont {Z.}~\bibnamefont
  {Kohley}},  \emph {et~al.},\ }\href@noop {} {\bibfield  {journal} {\bibinfo
  {journal} {Phys. Rev. C}\ }\textbf {\bibinfo {volume} {86}},\ \bibinfo
  {pages} {015803} (\bibinfo {year} {2012})}\BibitemShut {NoStop}%
\bibitem [{\citenamefont {Roca-Maza}\ \emph {et~al.}(2011)\citenamefont
  {Roca-Maza}, \citenamefont {Centelles}, \citenamefont {Vi\~nas},\ and\
  \citenamefont {Warda}}]{rocamaza11}%
  \BibitemOpen
  \bibfield  {author} {\bibinfo {author} {\bibfnamefont {X.}~\bibnamefont
  {Roca-Maza}}, \bibinfo {author} {\bibfnamefont {M.}~\bibnamefont
  {Centelles}}, \bibinfo {author} {\bibfnamefont {X.}~\bibnamefont {Vi\~nas}},
  \ and\ \bibinfo {author} {\bibfnamefont {M.}~\bibnamefont {Warda}},\ }\href
  {\doibase 10.1103/PhysRevLett.106.252501} {\bibfield  {journal} {\bibinfo
  {journal} {Phys. Rev. Lett.}\ }\textbf {\bibinfo {volume} {106}},\ \bibinfo
  {pages} {252501} (\bibinfo {year} {2011})}\BibitemShut {NoStop}%
\bibitem [{\citenamefont {Birkhan}\ \emph {et~al.}(2017)\citenamefont
  {Birkhan}, \citenamefont {Miorelli}, \citenamefont {Bacca}, \citenamefont
  {Bassauer}, \citenamefont {Bertulani}, \citenamefont {Hagen}, \citenamefont
  {Matsubara}, \citenamefont {von Neumann-Cosel}, \citenamefont {Papenbrock},
  \citenamefont {Pietralla} \emph {et~al.}}]{birkhan17}%
  \BibitemOpen
  \bibfield  {author} {\bibinfo {author} {\bibfnamefont {J.}~\bibnamefont
  {Birkhan}}, \bibinfo {author} {\bibfnamefont {M.}~\bibnamefont {Miorelli}},
  \bibinfo {author} {\bibfnamefont {S.}~\bibnamefont {Bacca}}, \bibinfo
  {author} {\bibfnamefont {S.}~\bibnamefont {Bassauer}}, \bibinfo {author}
  {\bibfnamefont {C.~A.}\ \bibnamefont {Bertulani}}, \bibinfo {author}
  {\bibfnamefont {G.}~\bibnamefont {Hagen}}, \bibinfo {author} {\bibfnamefont
  {H.}~\bibnamefont {Matsubara}}, \bibinfo {author} {\bibfnamefont
  {P.}~\bibnamefont {von Neumann-Cosel}}, \bibinfo {author} {\bibfnamefont
  {T.}~\bibnamefont {Papenbrock}}, \bibinfo {author} {\bibfnamefont
  {N.}~\bibnamefont {Pietralla}},  \emph {et~al.},\ }\href {\doibase
  10.1103/PhysRevLett.118.252501} {\bibfield  {journal} {\bibinfo  {journal}
  {Phys. Rev. Lett.}\ }\textbf {\bibinfo {volume} {118}},\ \bibinfo {pages}
  {252501} (\bibinfo {year} {2017})}\BibitemShut {NoStop}%
\bibitem [{\citenamefont {Hashimoto}\ \emph {et~al.}(2015)\citenamefont
  {Hashimoto}, \citenamefont {Krumbholz}, \citenamefont {Reinhard},
  \citenamefont {Tamii}, \citenamefont {von Neumann-Cosel}, \citenamefont
  {Adachi}, \citenamefont {Aoi}, \citenamefont {Bertulani}, \citenamefont
  {Fujita}, \citenamefont {Fujita} \emph {et~al.}}]{hashimoto15}%
  \BibitemOpen
  \bibfield  {author} {\bibinfo {author} {\bibfnamefont {T.}~\bibnamefont
  {Hashimoto}}, \bibinfo {author} {\bibfnamefont {A.~M.}\ \bibnamefont
  {Krumbholz}}, \bibinfo {author} {\bibfnamefont {P.-G.}\ \bibnamefont
  {Reinhard}}, \bibinfo {author} {\bibfnamefont {A.}~\bibnamefont {Tamii}},
  \bibinfo {author} {\bibfnamefont {P.}~\bibnamefont {von Neumann-Cosel}},
  \bibinfo {author} {\bibfnamefont {T.}~\bibnamefont {Adachi}}, \bibinfo
  {author} {\bibfnamefont {N.}~\bibnamefont {Aoi}}, \bibinfo {author}
  {\bibfnamefont {C.~A.}\ \bibnamefont {Bertulani}}, \bibinfo {author}
  {\bibfnamefont {H.}~\bibnamefont {Fujita}}, \bibinfo {author} {\bibfnamefont
  {Y.}~\bibnamefont {Fujita}},  \emph {et~al.},\ }\href {\doibase
  10.1103/PhysRevC.92.031305} {\bibfield  {journal} {\bibinfo  {journal} {Phys.
  Rev. C}\ }\textbf {\bibinfo {volume} {92}},\ \bibinfo {pages} {031305(R)}
  (\bibinfo {year} {2015})}\BibitemShut {NoStop}%
\bibitem [{\citenamefont {Tamii}\ \emph {et~al.}(2011)\citenamefont {Tamii},
  \citenamefont {Poltoratska}, \citenamefont {von Neumann-Cosel}, \citenamefont
  {Fujita}, \citenamefont {Adachi}, \citenamefont {Bertulani}, \citenamefont
  {Carter}, \citenamefont {Dozono}, \citenamefont {Fujita}, \citenamefont
  {Fujita} \emph {et~al.}}]{tamii11}%
  \BibitemOpen
  \bibfield  {author} {\bibinfo {author} {\bibfnamefont {A.}~\bibnamefont
  {Tamii}}, \bibinfo {author} {\bibfnamefont {I.}~\bibnamefont {Poltoratska}},
  \bibinfo {author} {\bibfnamefont {P.}~\bibnamefont {von Neumann-Cosel}},
  \bibinfo {author} {\bibfnamefont {Y.}~\bibnamefont {Fujita}}, \bibinfo
  {author} {\bibfnamefont {T.}~\bibnamefont {Adachi}}, \bibinfo {author}
  {\bibfnamefont {C.~A.}\ \bibnamefont {Bertulani}}, \bibinfo {author}
  {\bibfnamefont {J.}~\bibnamefont {Carter}}, \bibinfo {author} {\bibfnamefont
  {M.}~\bibnamefont {Dozono}}, \bibinfo {author} {\bibfnamefont
  {H.}~\bibnamefont {Fujita}}, \bibinfo {author} {\bibfnamefont
  {K.}~\bibnamefont {Fujita}},  \emph {et~al.},\ }\href {\doibase
  10.1103/PhysRevLett.107.062502} {\bibfield  {journal} {\bibinfo  {journal}
  {Phys. Rev. Lett.}\ }\textbf {\bibinfo {volume} {107}},\ \bibinfo {pages}
  {062502} (\bibinfo {year} {2011})}\BibitemShut {NoStop}%
\bibitem [{\citenamefont {Hagen}\ \emph {et~al.}(2015)\citenamefont {Hagen},
  \citenamefont {Ekstr\"om}, \citenamefont {Forss\'en}, \citenamefont {Jansen},
  \citenamefont {Nazarewicz}, \citenamefont {Papenbrock}, \citenamefont
  {Wendt}, \citenamefont {Bacca}, \citenamefont {Barnea}, \citenamefont
  {Carlsson} \emph {et~al.}}]{hagen15}%
  \BibitemOpen
  \bibfield  {author} {\bibinfo {author} {\bibfnamefont {G.}~\bibnamefont
  {Hagen}}, \bibinfo {author} {\bibfnamefont {A.}~\bibnamefont {Ekstr\"om}},
  \bibinfo {author} {\bibfnamefont {C.}~\bibnamefont {Forss\'en}}, \bibinfo
  {author} {\bibfnamefont {G.~R.}\ \bibnamefont {Jansen}}, \bibinfo {author}
  {\bibfnamefont {W.}~\bibnamefont {Nazarewicz}}, \bibinfo {author}
  {\bibfnamefont {T.}~\bibnamefont {Papenbrock}}, \bibinfo {author}
  {\bibfnamefont {K.~A.}\ \bibnamefont {Wendt}}, \bibinfo {author}
  {\bibfnamefont {S.}~\bibnamefont {Bacca}}, \bibinfo {author} {\bibfnamefont
  {N.}~\bibnamefont {Barnea}}, \bibinfo {author} {\bibfnamefont
  {B.}~\bibnamefont {Carlsson}},  \emph {et~al.},\ }\href {\doibase
  10.1038/nphys3529} {\bibfield  {journal} {\bibinfo  {journal} {Nat. Phys.}\
  }\textbf {\bibinfo {volume} {12}},\ \bibinfo {pages} {186} (\bibinfo {year}
  {2015})}\BibitemShut {NoStop}%
\bibitem [{\citenamefont {Epelbaum}\ \emph {et~al.}(2009)\citenamefont
  {Epelbaum}, \citenamefont {Hammer},\ and\ \citenamefont
  {Mei\ss{}ner}}]{epelbaum09}%
  \BibitemOpen
  \bibfield  {author} {\bibinfo {author} {\bibfnamefont {E.}~\bibnamefont
  {Epelbaum}}, \bibinfo {author} {\bibfnamefont {H.-W.}\ \bibnamefont
  {Hammer}}, \ and\ \bibinfo {author} {\bibfnamefont {U.-G.}\ \bibnamefont
  {Mei\ss{}ner}},\ }\href {\doibase 10.1103/RevModPhys.81.1773} {\bibfield
  {journal} {\bibinfo  {journal} {Rev. Mod. Phys.}\ }\textbf {\bibinfo {volume}
  {81}},\ \bibinfo {pages} {1773} (\bibinfo {year} {2009})}\BibitemShut
  {NoStop}%
\bibitem [{\citenamefont {Machleidt}\ and\ \citenamefont
  {Entem}(2011)}]{machleidt11}%
  \BibitemOpen
  \bibfield  {author} {\bibinfo {author} {\bibfnamefont {R.}~\bibnamefont
  {Machleidt}}\ and\ \bibinfo {author} {\bibfnamefont {D.}~\bibnamefont
  {Entem}},\ }\href {\doibase https://doi.org/10.1016/j.physrep.2011.02.001}
  {\bibfield  {journal} {\bibinfo  {journal} {Phys. Rep.}\ }\textbf {\bibinfo
  {volume} {503}},\ \bibinfo {pages} {1 } (\bibinfo {year} {2011})}\BibitemShut
  {NoStop}%
\bibitem [{\citenamefont {Hammer}\ \emph {et~al.}(2013)\citenamefont {Hammer},
  \citenamefont {Nogga},\ and\ \citenamefont {Schwenk}}]{Hamm13RMP}%
  \BibitemOpen
  \bibfield  {author} {\bibinfo {author} {\bibfnamefont {H.-W.}\ \bibnamefont
  {Hammer}}, \bibinfo {author} {\bibfnamefont {A.}~\bibnamefont {Nogga}}, \
  and\ \bibinfo {author} {\bibfnamefont {A.}~\bibnamefont {Schwenk}},\
  }\href@noop {} {\bibfield  {journal} {\bibinfo  {journal} {Rev. Mod. Phys.}\
  }\textbf {\bibinfo {volume} {85}},\ \bibinfo {pages} {197} (\bibinfo {year}
  {2013})}\BibitemShut {NoStop}%
\bibitem [{\citenamefont {Rossi}\ \emph {et~al.}(2013)\citenamefont {Rossi},
  \citenamefont {Adrich}, \citenamefont {Aksouh}, \citenamefont {Alvarez-Pol},
  \citenamefont {Aumann}, \citenamefont {Benlliure}, \citenamefont {B\"ohmer},
  \citenamefont {Boretzky}, \citenamefont {Casarejos}, \citenamefont {Chartier}
  \emph {et~al.}}]{rossi13}%
  \BibitemOpen
  \bibfield  {author} {\bibinfo {author} {\bibfnamefont {D.~M.}\ \bibnamefont
  {Rossi}}, \bibinfo {author} {\bibfnamefont {P.}~\bibnamefont {Adrich}},
  \bibinfo {author} {\bibfnamefont {F.}~\bibnamefont {Aksouh}}, \bibinfo
  {author} {\bibfnamefont {H.}~\bibnamefont {Alvarez-Pol}}, \bibinfo {author}
  {\bibfnamefont {T.}~\bibnamefont {Aumann}}, \bibinfo {author} {\bibfnamefont
  {J.}~\bibnamefont {Benlliure}}, \bibinfo {author} {\bibfnamefont
  {M.}~\bibnamefont {B\"ohmer}}, \bibinfo {author} {\bibfnamefont
  {K.}~\bibnamefont {Boretzky}}, \bibinfo {author} {\bibfnamefont
  {E.}~\bibnamefont {Casarejos}}, \bibinfo {author} {\bibfnamefont
  {M.}~\bibnamefont {Chartier}},  \emph {et~al.},\ }\href {\doibase
  10.1103/PhysRevLett.111.242503} {\bibfield  {journal} {\bibinfo  {journal}
  {Phys. Rev. Lett.}\ }\textbf {\bibinfo {volume} {111}},\ \bibinfo {pages}
  {242503} (\bibinfo {year} {2013})}\BibitemShut {NoStop}%
\bibitem [{\citenamefont {Marsh}(2014)}]{marsh14}%
  \BibitemOpen
  \bibfield  {author} {\bibinfo {author} {\bibfnamefont {B.~A.}\ \bibnamefont
  {Marsh}},\ }\href@noop {} {\bibfield  {journal} {\bibinfo  {journal} {Rev.
  Sci. Instrum.}\ }\textbf {\bibinfo {volume} {85}},\ \bibinfo {pages} {02B923}
  (\bibinfo {year} {2014})}\BibitemShut {NoStop}%
\bibitem [{\citenamefont {Fr{\r{a}}nberg}\ \emph {et~al.}(2008)\citenamefont
  {Fr{\r{a}}nberg}, \citenamefont {Delahaye}, \citenamefont {Billowes},
  \citenamefont {Blaum}, \citenamefont {Catherall}, \citenamefont {Duval},
  \citenamefont {Gianfrancesco}, \citenamefont {Giles}, \citenamefont
  {Jokinen}, \citenamefont {Lindroos} \emph {et~al.}}]{franberg08}%
  \BibitemOpen
  \bibfield  {author} {\bibinfo {author} {\bibfnamefont {H.}~\bibnamefont
  {Fr{\r{a}}nberg}}, \bibinfo {author} {\bibfnamefont {P.}~\bibnamefont
  {Delahaye}}, \bibinfo {author} {\bibfnamefont {J.}~\bibnamefont {Billowes}},
  \bibinfo {author} {\bibfnamefont {K.}~\bibnamefont {Blaum}}, \bibinfo
  {author} {\bibfnamefont {R.}~\bibnamefont {Catherall}}, \bibinfo {author}
  {\bibfnamefont {F.}~\bibnamefont {Duval}}, \bibinfo {author} {\bibfnamefont
  {O.}~\bibnamefont {Gianfrancesco}}, \bibinfo {author} {\bibfnamefont
  {T.}~\bibnamefont {Giles}}, \bibinfo {author} {\bibfnamefont
  {A.}~\bibnamefont {Jokinen}}, \bibinfo {author} {\bibfnamefont
  {M.}~\bibnamefont {Lindroos}},  \emph {et~al.},\ }\href {\doibase
  https://doi.org/10.1016/j.nimb.2008.05.097} {\bibfield  {journal} {\bibinfo
  {journal} {Nucl. Instr. Meth. Phys. Res. B}\ }\textbf {\bibinfo {volume}
  {266}},\ \bibinfo {pages} {4502 } (\bibinfo {year} {2008})}\BibitemShut
  {NoStop}%
\bibitem [{\citenamefont {Mueller}\ \emph {et~al.}(1983)\citenamefont
  {Mueller}, \citenamefont {Buchinger}, \citenamefont {Klempt}, \citenamefont
  {Otten}, \citenamefont {Neugart}, \citenamefont {Ekstr{\"o}m},\ and\
  \citenamefont {Heinemeier}}]{muller83}%
  \BibitemOpen
  \bibfield  {author} {\bibinfo {author} {\bibfnamefont {A.}~\bibnamefont
  {Mueller}}, \bibinfo {author} {\bibfnamefont {F.}~\bibnamefont {Buchinger}},
  \bibinfo {author} {\bibfnamefont {W.}~\bibnamefont {Klempt}}, \bibinfo
  {author} {\bibfnamefont {E.}~\bibnamefont {Otten}}, \bibinfo {author}
  {\bibfnamefont {R.}~\bibnamefont {Neugart}}, \bibinfo {author} {\bibfnamefont
  {C.}~\bibnamefont {Ekstr{\"o}m}}, \ and\ \bibinfo {author} {\bibfnamefont
  {J.}~\bibnamefont {Heinemeier}},\ }\href {\doibase
  https://doi.org/10.1016/0375-9474(83)90226-9} {\bibfield  {journal} {\bibinfo
   {journal} {Nucl. Phys. A}\ }\textbf {\bibinfo {volume} {403}},\ \bibinfo
  {pages} {234 } (\bibinfo {year} {1983})}\BibitemShut {NoStop}%
\bibitem [{\citenamefont {Klose}\ \emph {et~al.}(2012)\citenamefont {Klose},
  \citenamefont {Minamisono}, \citenamefont {Geppert}, \citenamefont
  {Fr{\"o}mmgen}, \citenamefont {Hammen}, \citenamefont {Kr{\"a}mer},
  \citenamefont {Krieger}, \citenamefont {Levy}, \citenamefont {Mantica},
  \citenamefont {N{\"o}rtersh{\"a}user} \emph {et~al.}}]{klose12}%
  \BibitemOpen
  \bibfield  {author} {\bibinfo {author} {\bibfnamefont {A.}~\bibnamefont
  {Klose}}, \bibinfo {author} {\bibfnamefont {K.}~\bibnamefont {Minamisono}},
  \bibinfo {author} {\bibfnamefont {C.}~\bibnamefont {Geppert}}, \bibinfo
  {author} {\bibfnamefont {N.}~\bibnamefont {Fr{\"o}mmgen}}, \bibinfo {author}
  {\bibfnamefont {M.}~\bibnamefont {Hammen}}, \bibinfo {author} {\bibfnamefont
  {J.}~\bibnamefont {Kr{\"a}mer}}, \bibinfo {author} {\bibfnamefont
  {A.}~\bibnamefont {Krieger}}, \bibinfo {author} {\bibfnamefont
  {C.}~\bibnamefont {Levy}}, \bibinfo {author} {\bibfnamefont {P.}~\bibnamefont
  {Mantica}}, \bibinfo {author} {\bibfnamefont {W.}~\bibnamefont
  {N{\"o}rtersh{\"a}user}},  \emph {et~al.},\ }\href {\doibase
  https://doi.org/10.1016/j.nima.2012.03.006} {\bibfield  {journal} {\bibinfo
  {journal} {Nucl. Instr. Meth. Phys. Res. A}\ }\textbf {\bibinfo {volume}
  {678}},\ \bibinfo {pages} {114 } (\bibinfo {year} {2012})}\BibitemShut
  {NoStop}%
\bibitem [{\citenamefont {Ryder}\ \emph {et~al.}(2015)\citenamefont {Ryder},
  \citenamefont {Minamisono}, \citenamefont {Asberry}, \citenamefont
  {Isherwood}, \citenamefont {Mantica}, \citenamefont {Miller}, \citenamefont
  {Rossi},\ and\ \citenamefont {Strum}}]{ryder15}%
  \BibitemOpen
  \bibfield  {author} {\bibinfo {author} {\bibfnamefont {C.}~\bibnamefont
  {Ryder}}, \bibinfo {author} {\bibfnamefont {K.}~\bibnamefont {Minamisono}},
  \bibinfo {author} {\bibfnamefont {H.}~\bibnamefont {Asberry}}, \bibinfo
  {author} {\bibfnamefont {B.}~\bibnamefont {Isherwood}}, \bibinfo {author}
  {\bibfnamefont {P.}~\bibnamefont {Mantica}}, \bibinfo {author} {\bibfnamefont
  {A.}~\bibnamefont {Miller}}, \bibinfo {author} {\bibfnamefont
  {D.}~\bibnamefont {Rossi}}, \ and\ \bibinfo {author} {\bibfnamefont
  {R.}~\bibnamefont {Strum}},\ }\href {\doibase
  https://doi.org/10.1016/j.sab.2015.08.004} {\bibfield  {journal} {\bibinfo
  {journal} {Spectroch. Acta B}\ }\textbf {\bibinfo {volume} {113}},\ \bibinfo
  {pages} {16 } (\bibinfo {year} {2015})}\BibitemShut {NoStop}%
\bibitem [{\citenamefont {Hammen}\ \emph {et~al.}(2018)\citenamefont {Hammen},
  \citenamefont {N\"ortersh\"auser}, \citenamefont {Balabanski}, \citenamefont
  {Bissell}, \citenamefont {Blaum}, \citenamefont {Budin\ifmmode \check{c}\else
  \v{c}\fi{}evi\ifmmode~\acute{c}\else \'{c}\fi{}}, \citenamefont {Cheal},
  \citenamefont {Flanagan}, \citenamefont {Fr\"ommgen}, \citenamefont
  {Georgiev} \emph {et~al.}}]{hammen18}%
  \BibitemOpen
  \bibfield  {author} {\bibinfo {author} {\bibfnamefont {M.}~\bibnamefont
  {Hammen}}, \bibinfo {author} {\bibfnamefont {W.}~\bibnamefont
  {N\"ortersh\"auser}}, \bibinfo {author} {\bibfnamefont {D.~L.}\ \bibnamefont
  {Balabanski}}, \bibinfo {author} {\bibfnamefont {M.~L.}\ \bibnamefont
  {Bissell}}, \bibinfo {author} {\bibfnamefont {K.}~\bibnamefont {Blaum}},
  \bibinfo {author} {\bibfnamefont {I.}~\bibnamefont {Budin\ifmmode
  \check{c}\else \v{c}\fi{}evi\ifmmode~\acute{c}\else \'{c}\fi{}}}, \bibinfo
  {author} {\bibfnamefont {B.}~\bibnamefont {Cheal}}, \bibinfo {author}
  {\bibfnamefont {K.~T.}\ \bibnamefont {Flanagan}}, \bibinfo {author}
  {\bibfnamefont {N.}~\bibnamefont {Fr\"ommgen}}, \bibinfo {author}
  {\bibfnamefont {G.}~\bibnamefont {Georgiev}},  \emph {et~al.},\ }\href
  {\doibase 10.1103/PhysRevLett.121.102501} {\bibfield  {journal} {\bibinfo
  {journal} {Phys. Rev. Lett.}\ }\textbf {\bibinfo {volume} {121}},\ \bibinfo
  {pages} {102501} (\bibinfo {year} {2018})}\BibitemShut {NoStop}%
\bibitem [{\citenamefont {Fricke}\ and\ \citenamefont
  {Heilig}(2004)}]{fricke04}%
  \BibitemOpen
  \bibfield  {author} {\bibinfo {author} {\bibfnamefont {G.}~\bibnamefont
  {Fricke}}\ and\ \bibinfo {author} {\bibfnamefont {K.}~\bibnamefont
  {Heilig}},\ }\href@noop {} {\emph {\bibinfo {title} {{Nuclear Charge
  Radii}}}},\ \bibinfo {series} {Group~I: Elementary Particles, Nuclei and
  Atoms}, Vol.~\bibinfo {volume} {20}\ (\bibinfo  {publisher} {Springer},\
  \bibinfo {year} {2004})\BibitemShut {NoStop}%
\bibitem [{\citenamefont {Miorelli}\ \emph {et~al.}(2018)\citenamefont
  {Miorelli}, \citenamefont {Bacca}, \citenamefont {Hagen},\ and\ \citenamefont
  {Papenbrock}}]{miorelli18}%
  \BibitemOpen
  \bibfield  {author} {\bibinfo {author} {\bibfnamefont {M.}~\bibnamefont
  {Miorelli}}, \bibinfo {author} {\bibfnamefont {S.}~\bibnamefont {Bacca}},
  \bibinfo {author} {\bibfnamefont {G.}~\bibnamefont {Hagen}}, \ and\ \bibinfo
  {author} {\bibfnamefont {T.}~\bibnamefont {Papenbrock}},\ }\href {\doibase
  10.1103/PhysRevC.98.014324} {\bibfield  {journal} {\bibinfo  {journal} {Phys.
  Rev. C}\ }\textbf {\bibinfo {volume} {98}},\ \bibinfo {pages} {014324}
  (\bibinfo {year} {2018})}\BibitemShut {NoStop}%
\bibitem [{\citenamefont {Simonis}\ \emph {et~al.}(2019)\citenamefont
  {Simonis}, \citenamefont {Bacca},\ and\ \citenamefont {Hagen}}]{Simonis}%
  \BibitemOpen
  \bibfield  {author} {\bibinfo {author} {\bibfnamefont {J.}~\bibnamefont
  {Simonis}}, \bibinfo {author} {\bibfnamefont {S.}~\bibnamefont {Bacca}}, \
  and\ \bibinfo {author} {\bibfnamefont {G.}~\bibnamefont {Hagen}},\ }\href
  {\doibase 10.1140/epja/i2019-12825-0} {\bibfield  {journal} {\bibinfo
  {journal} {Eur. Phys. J. A}\ }\textbf {\bibinfo {volume} {55}},\ \bibinfo
  {pages} {241} (\bibinfo {year} {2019})}\BibitemShut {NoStop}%
\bibitem [{\citenamefont {Bacca}\ \emph {et~al.}(2013)\citenamefont {Bacca},
  \citenamefont {Barnea}, \citenamefont {Hagen}, \citenamefont {Orlandini},\
  and\ \citenamefont {Papenbrock}}]{bacca2013}%
  \BibitemOpen
  \bibfield  {author} {\bibinfo {author} {\bibfnamefont {S.}~\bibnamefont
  {Bacca}}, \bibinfo {author} {\bibfnamefont {N.}~\bibnamefont {Barnea}},
  \bibinfo {author} {\bibfnamefont {G.}~\bibnamefont {Hagen}}, \bibinfo
  {author} {\bibfnamefont {G.}~\bibnamefont {Orlandini}}, \ and\ \bibinfo
  {author} {\bibfnamefont {T.}~\bibnamefont {Papenbrock}},\ }\href {\doibase
  10.1103/PhysRevLett.111.122502} {\bibfield  {journal} {\bibinfo  {journal}
  {Phys. Rev. Lett.}\ }\textbf {\bibinfo {volume} {111}},\ \bibinfo {pages}
  {122502} (\bibinfo {year} {2013})}\BibitemShut {NoStop}%
\bibitem [{\citenamefont {Bacca}\ \emph {et~al.}(2014)\citenamefont {Bacca},
  \citenamefont {Barnea}, \citenamefont {Hagen}, \citenamefont {Miorelli},
  \citenamefont {Orlandini},\ and\ \citenamefont {Papenbrock}}]{bacca2014}%
  \BibitemOpen
  \bibfield  {author} {\bibinfo {author} {\bibfnamefont {S.}~\bibnamefont
  {Bacca}}, \bibinfo {author} {\bibfnamefont {N.}~\bibnamefont {Barnea}},
  \bibinfo {author} {\bibfnamefont {G.}~\bibnamefont {Hagen}}, \bibinfo
  {author} {\bibfnamefont {M.}~\bibnamefont {Miorelli}}, \bibinfo {author}
  {\bibfnamefont {G.}~\bibnamefont {Orlandini}}, \ and\ \bibinfo {author}
  {\bibfnamefont {T.}~\bibnamefont {Papenbrock}},\ }\href {\doibase
  10.1103/PhysRevC.90.064619} {\bibfield  {journal} {\bibinfo  {journal} {Phys.
  Rev. C}\ }\textbf {\bibinfo {volume} {90}},\ \bibinfo {pages} {064619}
  (\bibinfo {year} {2014})}\BibitemShut {NoStop}%
\bibitem [{\citenamefont {Miorelli}\ \emph {et~al.}(2016)\citenamefont
  {Miorelli}, \citenamefont {Bacca}, \citenamefont {Barnea}, \citenamefont
  {Hagen}, \citenamefont {Jansen}, \citenamefont {Orlandini},\ and\
  \citenamefont {Papenbrock}}]{miorelli2016}%
  \BibitemOpen
  \bibfield  {author} {\bibinfo {author} {\bibfnamefont {M.}~\bibnamefont
  {Miorelli}}, \bibinfo {author} {\bibfnamefont {S.}~\bibnamefont {Bacca}},
  \bibinfo {author} {\bibfnamefont {N.}~\bibnamefont {Barnea}}, \bibinfo
  {author} {\bibfnamefont {G.}~\bibnamefont {Hagen}}, \bibinfo {author}
  {\bibfnamefont {G.~R.}\ \bibnamefont {Jansen}}, \bibinfo {author}
  {\bibfnamefont {G.}~\bibnamefont {Orlandini}}, \ and\ \bibinfo {author}
  {\bibfnamefont {T.}~\bibnamefont {Papenbrock}},\ }\href {\doibase
  10.1103/PhysRevC.94.034317} {\bibfield  {journal} {\bibinfo  {journal} {Phys.
  Rev. C}\ }\textbf {\bibinfo {volume} {94}},\ \bibinfo {pages} {034317}
  (\bibinfo {year} {2016})}\BibitemShut {NoStop}%
\bibitem [{\citenamefont {Roca-Maza}\ \emph {et~al.}(2015)\citenamefont
  {Roca-Maza}, \citenamefont {Vi\~nas}, \citenamefont {Centelles},
  \citenamefont {Agrawal}, \citenamefont {Col\`o}, \citenamefont {Paar},
  \citenamefont {Piekarewicz},\ and\ \citenamefont {Vretenar}}]{rocamaza15}%
  \BibitemOpen
  \bibfield  {author} {\bibinfo {author} {\bibfnamefont {X.}~\bibnamefont
  {Roca-Maza}}, \bibinfo {author} {\bibfnamefont {X.}~\bibnamefont {Vi\~nas}},
  \bibinfo {author} {\bibfnamefont {M.}~\bibnamefont {Centelles}}, \bibinfo
  {author} {\bibfnamefont {B.~K.}\ \bibnamefont {Agrawal}}, \bibinfo {author}
  {\bibfnamefont {G.}~\bibnamefont {Col\`o}}, \bibinfo {author} {\bibfnamefont
  {N.}~\bibnamefont {Paar}}, \bibinfo {author} {\bibfnamefont {J.}~\bibnamefont
  {Piekarewicz}}, \ and\ \bibinfo {author} {\bibfnamefont {D.}~\bibnamefont
  {Vretenar}},\ }\href {\doibase 10.1103/PhysRevC.92.064304} {\bibfield
  {journal} {\bibinfo  {journal} {Phys. Rev. C}\ }\textbf {\bibinfo {volume}
  {92}},\ \bibinfo {pages} {064304} (\bibinfo {year} {2015})}\BibitemShut
  {NoStop}%
\bibitem [{\citenamefont {Hebeler}\ \emph {et~al.}(2011)\citenamefont
  {Hebeler}, \citenamefont {Bogner}, \citenamefont {Furnstahl}, \citenamefont
  {Nogga},\ and\ \citenamefont {Schwenk}}]{Hebeler2011}%
  \BibitemOpen
  \bibfield  {author} {\bibinfo {author} {\bibfnamefont {K.}~\bibnamefont
  {Hebeler}}, \bibinfo {author} {\bibfnamefont {S.~K.}\ \bibnamefont {Bogner}},
  \bibinfo {author} {\bibfnamefont {R.~J.}\ \bibnamefont {Furnstahl}}, \bibinfo
  {author} {\bibfnamefont {A.}~\bibnamefont {Nogga}}, \ and\ \bibinfo {author}
  {\bibfnamefont {A.}~\bibnamefont {Schwenk}},\ }\href {\doibase
  10.1103/PhysRevC.83.031301} {\bibfield  {journal} {\bibinfo  {journal} {Phys.
  Rev. C}\ }\textbf {\bibinfo {volume} {83}},\ \bibinfo {pages} {031301(R)}
  (\bibinfo {year} {2011})}\BibitemShut {NoStop}%
\bibitem [{\citenamefont {Ekstr\"om}\ \emph {et~al.}(2015)\citenamefont
  {Ekstr\"om}, \citenamefont {Jansen}, \citenamefont {Wendt}, \citenamefont
  {Hagen}, \citenamefont {Papenbrock}, \citenamefont {Carlsson}, \citenamefont
  {Forss\'en}, \citenamefont {Hjorth-Jensen}, \citenamefont {Navr\'atil},\ and\
  \citenamefont {Nazarewicz}}]{ekstrom2015}%
  \BibitemOpen
  \bibfield  {author} {\bibinfo {author} {\bibfnamefont {A.}~\bibnamefont
  {Ekstr\"om}}, \bibinfo {author} {\bibfnamefont {G.~R.}\ \bibnamefont
  {Jansen}}, \bibinfo {author} {\bibfnamefont {K.~A.}\ \bibnamefont {Wendt}},
  \bibinfo {author} {\bibfnamefont {G.}~\bibnamefont {Hagen}}, \bibinfo
  {author} {\bibfnamefont {T.}~\bibnamefont {Papenbrock}}, \bibinfo {author}
  {\bibfnamefont {B.~D.}\ \bibnamefont {Carlsson}}, \bibinfo {author}
  {\bibfnamefont {C.}~\bibnamefont {Forss\'en}}, \bibinfo {author}
  {\bibfnamefont {M.}~\bibnamefont {Hjorth-Jensen}}, \bibinfo {author}
  {\bibfnamefont {P.}~\bibnamefont {Navr\'atil}}, \ and\ \bibinfo {author}
  {\bibfnamefont {W.}~\bibnamefont {Nazarewicz}},\ }\href {\doibase
  10.1103/PhysRevC.91.051301} {\bibfield  {journal} {\bibinfo  {journal} {Phys.
  Rev. C}\ }\textbf {\bibinfo {volume} {91}},\ \bibinfo {pages} {051301(R)}
  (\bibinfo {year} {2015})}\BibitemShut {NoStop}%
\bibitem [{\citenamefont {Simonis}\ \emph {et~al.}(2016)\citenamefont
  {Simonis}, \citenamefont {Hebeler}, \citenamefont {Holt}, \citenamefont
  {Men{\'e}ndez},\ and\ \citenamefont {Schwenk}}]{Simo16unc}%
  \BibitemOpen
  \bibfield  {author} {\bibinfo {author} {\bibfnamefont {J.}~\bibnamefont
  {Simonis}}, \bibinfo {author} {\bibfnamefont {K.}~\bibnamefont {Hebeler}},
  \bibinfo {author} {\bibfnamefont {J.~D.}\ \bibnamefont {Holt}}, \bibinfo
  {author} {\bibfnamefont {J.}~\bibnamefont {Men{\'e}ndez}}, \ and\ \bibinfo
  {author} {\bibfnamefont {A.}~\bibnamefont {Schwenk}},\ }\href {\doibase
  10.1103/PhysRevC.93.011302} {\bibfield  {journal} {\bibinfo  {journal} {Phys.
  Rev. C}\ }\textbf {\bibinfo {volume} {93}},\ \bibinfo {pages} {011302(R)}
  (\bibinfo {year} {2016})}\BibitemShut {NoStop}%
\bibitem [{\citenamefont {Simonis}\ \emph {et~al.}(2017)\citenamefont
  {Simonis}, \citenamefont {Stroberg}, \citenamefont {Hebeler}, \citenamefont
  {Holt},\ and\ \citenamefont {Schwenk}}]{Simo17SatFinNuc}%
  \BibitemOpen
  \bibfield  {author} {\bibinfo {author} {\bibfnamefont {J.}~\bibnamefont
  {Simonis}}, \bibinfo {author} {\bibfnamefont {S.~R.}\ \bibnamefont
  {Stroberg}}, \bibinfo {author} {\bibfnamefont {K.}~\bibnamefont {Hebeler}},
  \bibinfo {author} {\bibfnamefont {J.~D.}\ \bibnamefont {Holt}}, \ and\
  \bibinfo {author} {\bibfnamefont {A.}~\bibnamefont {Schwenk}},\ }\href
  {\doibase 10.1103/PhysRevC.96.014303} {\bibfield  {journal} {\bibinfo
  {journal} {Phys. Rev. C}\ }\textbf {\bibinfo {volume} {96}},\ \bibinfo
  {pages} {014303} (\bibinfo {year} {2017})}\BibitemShut {NoStop}%
\bibitem [{\citenamefont {Morris}\ \emph {et~al.}(2018)\citenamefont {Morris},
  \citenamefont {Simonis}, \citenamefont {Stroberg}, \citenamefont {Stumpf},
  \citenamefont {Hagen}, \citenamefont {Holt}, \citenamefont {Jansen},
  \citenamefont {Papenbrock}, \citenamefont {Roth},\ and\ \citenamefont
  {Schwenk}}]{Morr17Tin}%
  \BibitemOpen
  \bibfield  {author} {\bibinfo {author} {\bibfnamefont {T.~D.}\ \bibnamefont
  {Morris}}, \bibinfo {author} {\bibfnamefont {J.}~\bibnamefont {Simonis}},
  \bibinfo {author} {\bibfnamefont {S.~R.}\ \bibnamefont {Stroberg}}, \bibinfo
  {author} {\bibfnamefont {C.}~\bibnamefont {Stumpf}}, \bibinfo {author}
  {\bibfnamefont {G.}~\bibnamefont {Hagen}}, \bibinfo {author} {\bibfnamefont
  {J.~D.}\ \bibnamefont {Holt}}, \bibinfo {author} {\bibfnamefont {G.~R.}\
  \bibnamefont {Jansen}}, \bibinfo {author} {\bibfnamefont {T.}~\bibnamefont
  {Papenbrock}}, \bibinfo {author} {\bibfnamefont {R.}~\bibnamefont {Roth}}, \
  and\ \bibinfo {author} {\bibfnamefont {A.}~\bibnamefont {Schwenk}},\ }\href
  {\doibase 10.1103/PhysRevLett.120.152503} {\bibfield  {journal} {\bibinfo
  {journal} {Phys. Rev. Lett.}\ }\textbf {\bibinfo {volume} {120}},\ \bibinfo
  {pages} {152503} (\bibinfo {year} {2018})}\BibitemShut {NoStop}%
\bibitem [{\citenamefont {Holt}\ \emph {et~al.}(2019)\citenamefont {Holt},
  \citenamefont {Stroberg}, \citenamefont {Schwenk},\ and\ \citenamefont
  {Simonis}}]{Holt2019}%
  \BibitemOpen
  \bibfield  {author} {\bibinfo {author} {\bibfnamefont {J.~D.}\ \bibnamefont
  {Holt}}, \bibinfo {author} {\bibfnamefont {S.~R.}\ \bibnamefont {Stroberg}},
  \bibinfo {author} {\bibfnamefont {A.}~\bibnamefont {Schwenk}}, \ and\
  \bibinfo {author} {\bibfnamefont {J.}~\bibnamefont {Simonis}},\ }\href@noop
  {} {\  (\bibinfo {year} {2019})},\ \Eprint {http://arxiv.org/abs/1905.10475}
  {arXiv:1905.10475} \BibitemShut {NoStop}%
\bibitem [{cod(2019)}]{codata18}%
  \BibitemOpen
  \href@noop {} {}\bibinfo {howpublished}
  {\url{https://physics.nist.gov/cgi-bin/cuu/Value?rp}} (\bibinfo {year}
  {2019})\BibitemShut {NoStop}%
\bibitem [{\citenamefont {Kopecky}\ \emph {et~al.}(1997)\citenamefont
  {Kopecky}, \citenamefont {Harvey}, \citenamefont {Hill}, \citenamefont
  {Krenn}, \citenamefont {Pernicka}, \citenamefont {Riehs},\ and\ \citenamefont
  {Steiner}}]{kopecky97}%
  \BibitemOpen
  \bibfield  {author} {\bibinfo {author} {\bibfnamefont {S.}~\bibnamefont
  {Kopecky}}, \bibinfo {author} {\bibfnamefont {J.~A.}\ \bibnamefont {Harvey}},
  \bibinfo {author} {\bibfnamefont {N.~W.}\ \bibnamefont {Hill}}, \bibinfo
  {author} {\bibfnamefont {M.}~\bibnamefont {Krenn}}, \bibinfo {author}
  {\bibfnamefont {M.}~\bibnamefont {Pernicka}}, \bibinfo {author}
  {\bibfnamefont {P.}~\bibnamefont {Riehs}}, \ and\ \bibinfo {author}
  {\bibfnamefont {S.}~\bibnamefont {Steiner}},\ }\href {\doibase
  10.1103/PhysRevC.56.2229} {\bibfield  {journal} {\bibinfo  {journal} {Phys.
  Rev. C}\ }\textbf {\bibinfo {volume} {56}},\ \bibinfo {pages} {2229}
  (\bibinfo {year} {1997})}\BibitemShut {NoStop}%
\bibitem [{\citenamefont {Friar}\ \emph {et~al.}(1997)\citenamefont {Friar},
  \citenamefont {Martorell},\ and\ \citenamefont {Sprung}}]{friar97}%
  \BibitemOpen
  \bibfield  {author} {\bibinfo {author} {\bibfnamefont {J.~L.}\ \bibnamefont
  {Friar}}, \bibinfo {author} {\bibfnamefont {J.}~\bibnamefont {Martorell}}, \
  and\ \bibinfo {author} {\bibfnamefont {D.~W.~L.}\ \bibnamefont {Sprung}},\
  }\href {\doibase 10.1103/PhysRevA.56.4579} {\bibfield  {journal} {\bibinfo
  {journal} {Phys. Rev. A}\ }\textbf {\bibinfo {volume} {56}},\ \bibinfo
  {pages} {4579} (\bibinfo {year} {1997})}\BibitemShut {NoStop}%
\bibitem [{\citenamefont {Raimondi}\ and\ \citenamefont
  {Barbieri}(2019)}]{Raimondi:2018mtv}%
  \BibitemOpen
  \bibfield  {author} {\bibinfo {author} {\bibfnamefont {F.}~\bibnamefont
  {Raimondi}}\ and\ \bibinfo {author} {\bibfnamefont {C.}~\bibnamefont
  {Barbieri}},\ }\href {\doibase 10.1103/PhysRevC.99.054327} {\bibfield
  {journal} {\bibinfo  {journal} {Phys. Rev. C}\ }\textbf {\bibinfo {volume}
  {99}},\ \bibinfo {pages} {054327} (\bibinfo {year} {2019})}\BibitemShut
  {NoStop}%
\end{thebibliography}
%

\end{document}